\documentclass[conference, a4paper]{IEEEtran}
\IEEEoverridecommandlockouts
\usepackage{cite}
\usepackage[pdftex]{graphicx}
\usepackage{caption}
\usepackage{epstopdf}
\usepackage{dblfloatfix}
\usepackage{blindtext}
\usepackage{float}
\usepackage[cmex10]{amsmath}
\usepackage{algorithm} 
\usepackage{algpseudocode} 
\usepackage{amssymb}
\usepackage{color}
\usepackage{bm}
\usepackage{amsfonts}
\usepackage{amsmath}
\usepackage{amsmath,mathtools,amssymb}
\usepackage{mathrsfs}
\usepackage[mathlines]{lineno}
\usepackage[utf8]{inputenc}
\usepackage[english]{babel}
\usepackage{bookmark}
\DeclareMathOperator{\sign}{sign}
\DeclarePairedDelimiter{\norm}{\lVert}{\rVert}
\usepackage{stmaryrd,xspace}
\newcommand{\normmm}[1]{{\left\vert\kern-0.25ex\left\vert\kern-0.25ex\left\vert #1 
   \right\vert\kern-0.25ex\right\vert\kern-0.25ex\right\vert}}
\renewcommand{\figurename}{Fig.}

\begin{document}

\title{Sparse Channel Reconstruction With Nonconvex Regularizer via DC Programming for Massive MIMO Systems} 

\author{\IEEEauthorblockN{Pengxia Wu\IEEEauthorrefmark{1}, Hui Ma\IEEEauthorrefmark{1} and Julian Cheng\IEEEauthorrefmark{1}}\\
\IEEEauthorblockA{\IEEEauthorrefmark{1}School of Engineering, The University of British Columbia, Kelowna, BC, Canada\\
Email: {pengxia.wu@ubc.ca}, {hui.ma@alumni.ubc.ca}, {julian.cheng@ubc.ca}}

 \thanks{{Accepted in 2020 IEEE Global Communications Conference.}}
}
\maketitle

\begin{abstract}

	Sparse channel estimation for massive multiple-input multiple-output systems has drawn much attention in recent years. The required pilots are substantially reduced when the sparse channel state vectors can be reconstructed from a few numbers of measurements. A popular approach for sparse reconstruction is to solve the least-squares problem with a convex regularization. However, the convex regularizer is either too loose to force sparsity or lead to biased estimation. In this paper, the sparse channel reconstruction is solved by minimizing the least-squares objective with a nonconvex regularizer, which can exactly express the sparsity constraint and avoid introducing serious bias in the solution. A novel algorithm is proposed for solving the resulting nonconvex optimization via the difference of convex functions programming and the gradient projection descent. Simulation results show that the proposed algorithm is fast and accurate, and it outperforms the existing sparse recovery algorithms in terms of reconstruction errors.
	
\end{abstract}


\IEEEpeerreviewmaketitle
\section{Introduction}
	 Channel estimation for massive multiple-input multiple-output (MIMO) systems has drawn increasing attention in recent years.
	 Due to the large-scale antenna arrays, conventional channel estimation techniques require a large number of time slots as the pilot training overhead.
	 Sparse channel estimation \cite{bajwa2010compressed, rao2014distributed,gao2015spatially, gao2016structured, gao2017reliable} has the advantage to reduce pilot training overhead substantially by exploiting the channel sparsity in angular domain.
	Sparse channel estimation requires to reconstruct a sparse channel state vector from a few number of randomly projected measurements. 
	Thus, a sparse channel estimation process also refers to a sparse channel reconstruction.
	A sparse channel reconstruction is to resolve an underdetermined linear system of equations to achieve a sparse solution closest to the real angular-domain channel vector.
	A unique sparse solution can be obtained through solving an optimization of least squares constrained with a \mbox{$\ell_0$-norm} term, where the \mbox{$\ell_0$-norm} constraint term is used to enforce sparsity in the solution by limiting the number of \mbox{non-zero} elements.
	Since \mbox{$\ell_0$-norm} is discrete and nonconvex, the ultimate sparse reconstruction problem is a combinational optimization problem and is \mbox{NP-hard} \cite{amaldi1998on}.
	
	Common approaches to solving sparse reconstruction include greedy approach and $\ell_1$-relaxation optimization.
	Representative greedy algorithms include the orthogonal matching pursuit (OMP) \cite{pati1993orthogonal, tropp2007signal}, the CoSaMP \cite{needell2009CoSaMP} and the least angle regression (LARS) \cite{efron2004least}.
	These methods work well when the vector is sufficiently sparse, but their performances degrade seriously when the sparsity is reduced. 
	The most popular approach for sparse reconstruction is the \mbox{$\ell_1$-relaxation} optimization, where the nonconvex $\ell_0$-norm constraint is relaxed and approximated by the convex \mbox{$\ell_1$-norm} constraint.
	 Numerous algorithms have been developed for solving the relaxed convex optimization of sparse reconstructions, such as the iterative shrinkage-thresholding algorithm (ISTA) \cite{blumensath2008thomas}, the fast iterative shrinkage-thresholding algorithm (FISTA) \cite{Beck2009Afast}, the gradient projection sparse recovery algorithm (GPSR) \cite{figueiredo2007gradient} and the sparse reconstruction by separable approximation (SpaRSA) \cite{wright2009sparse}.
	These \mbox{$\ell_1$-relaxation} based algorithms can guarantee to converge theoretically within a finite number of iterations.
	However, the $\ell_1$-norm constraint is a loose relaxation of the $\ell_0$-norm constraint, and does not always provide an accurate sparse solution.

	In this paper, we propose a novel algorithm for sparse channel reconstruction for massive MIMO systems.
	More specifically, we introduce the top-$(K, 1)$ norm \cite{Gotoh2018DC} to represent exactly the $\ell_0$-norm constraint instead of its approximation. 
	Then, we express the problem of sparse reconstructions as the optimization of least squares objective penalised by a nonconvex regularizer term, which is represented using the \mbox{top-$(K, 1)$} norm.
	To solve the resulting nonconvex optimization problem of sparse reconstructions, we employ the difference of convex functions (DC) programming \cite{Tao1997convex}.
	In specific, the DC programming solves nonconvex optimization by decomposing the nonconvex objective into a form of DC and performing iterations through a primal-dual method \cite{Tao1997convex}. 
	At each iteration, the DC algorithm solves a convex subproblem, which is an approximation of the original nonconvex problem.
	We express the subproblem as a bound-constrained quadratic program (BCQP) with a simple nonnegativity constraint, such that it can be efficiently solved by the gradient projection descent method.
	The proposed DC gradient projection algorithm is a double-layer iteration algorithm, which shares the same global convergence property with general DC algorithms \cite{Tao1997convex}.
	Numerical results show that the proposed DC gradient projection algorithm can accurately reconstruct the sparse channel vectors both in the noiseless and noisy scenario.
	The proposed DC gradient projection algorithm can achieve solutions that have lower reconstruction errors compared to existing algorithms, including the OMP algorithm and several \mbox{$\ell_1$-relaxation} algorithms.

\section{System Model}
	We consider a downlink massive MIMO system, where the base station (BS) has $N$ antennas and each user equipment (UE) has a single antenna.
	 We let the vector $\mathbf h_s \in \mathbb{C}^{N}$ denote the \mbox{spatial-domain} channel between the BS and a UE, and let the vector $\mathbf h_a \in \mathbb{C}^{N}$ denote channel in virtual angular-domain.	
	Assuming a narrowband blockfading channel, the spatial-domain channel vector $\mathbf h_s$ is given by \cite{tse2005fundamentals} 
	\begin{IEEEeqnarray*}{lCl}\label{spatial channel vector}
	\mathbf h_s = \sqrt{\frac{N}{N_p}} \sum _{l=1} ^{N_p} \beta^{(l)} \bm \alpha (\phi ^{(l)}) \IEEEyesnumber
	\end{IEEEeqnarray*}
    where $N_p$ is the number of paths; $l = 1$ is the index for the line-of-sight path; $2 \leq l \leq N_p$ is the index for non-line-of-sight paths; $\beta^{(l)}$ is the complex path gain;
    $\bm \alpha (\phi  ^{(l)})$ is the corresponding array steering vector that contains a list of complex spatial sinusoids to represent the relative phase shifts for the incident far-field waveform across the array elements.
	For the $N$-element uniform linear array, the array steering vector $\bm \alpha (\phi  ^{(l)})$ is given by
    \begin{IEEEeqnarray*}{lCl}
    \label{steering vector}
    \bm \alpha (\phi ^{(l)}) = \frac{1}{\sqrt{N}} [1,  e^{-j2\pi \phi^{(l)}}, ... , e^{-j2\pi \phi^{(l)} (N-1)}]^\mathrm{T}
    \IEEEyesnumber
    \end{IEEEeqnarray*}	 
	where $\phi ^{(l)}$ denotes the spatial direction of the $l$th path, and it is related to the physical angle $\theta^{(l)}$ by $\phi ^{(l)} = \frac{d}{\lambda} \sin{\theta ^{(l)}}$ for $-\frac{1}{2} \leq \phi ^{(l)} \leq \frac{1}{2}$ and $-\frac{\pi}{2} \leq \theta ^{(l)} \leq \frac{\pi}{2}$, where $\lambda$ is the wavelength, and $d =\frac{\lambda}{2}$ is the antenna spacing. 

	The spatial-domain channel vector $\mathbf h_s$ in \eqref{spatial channel vector} can be transformed into the virtual angular-domain by \cite{tse2005fundamentals}
	\begin{IEEEeqnarray*}{lCl} \label{beamspace_channel_vector}
	\mathbf h_a = \mathbf{U} \mathbf h_s 
	\IEEEyesnumber
	\end{IEEEeqnarray*}    
	where $\mathbf U$ denotes the discrete Fourier transform (DFT) matrix having the size $N \times N$, and it can be expressed using a set of orthogonal array steering vectors as 
	\begin{IEEEeqnarray*}{lCl}
	\label{U_matrix}
	\mathbf{U}  = [\bm \alpha(\phi_1),  \bm \alpha(\phi_2),  ...  , \bm \alpha(\phi_N)]^H
	\IEEEyesnumber
	\end{IEEEeqnarray*}  
   where $\phi_i = \frac{1}{N} (i-\frac{N+1}{2})$ for $i=1,2,...,N$ is the spatial direction predefined by the array having half-wavelength spaced antennas.
    The massive MIMO channels have strong spatially correlations and much-lower degrees of freedom than the number of antennas.
   Since the BS is usually located at an elevated position far away from the UE, the number of scattering clusters around the BS is limited and each scattering cluster has a small angular spread.
   Thus, the majority of channel energy $\norm{\mathbf h_a}_2^2$ is occupied by a limited dimensions in angular domain.
   Consequently, the massive MIMO channels have a sparse angular-domain representation, and only a small number of nonzero elements exist in the angular-domain channel vector $\mathbf h_a$.

	For the pilot-aided downlink channel estimation schemes, the BS transmits the known pilots $\mathbf{P}$ to the UEs. The received pilot symbols at the UE can be expressed as \cite{choi2017compressed} 
    \begin{IEEEeqnarray*}{lCl}
    \label{received_pilots}
    \mathbf{r} = \mathbf{P}\mathbf{h}_s + \mathbf{w} 
    \IEEEyesnumber
    \end{IEEEeqnarray*}
	where $\mathbf{r} \in \mathbb{C}^{M}$ is the received pilots; $\mathbf P \in \mathbb{C}^{M \times N}$ is the downlink pilot matrix transmitted over $M$ time slots; $\mathbf{h}_s \in \mathbb{C}^{N}$ is the spatial-domain channel vector; $\mathbf{w} $ is the received noise vector and $\mathbf{w} \sim \mathcal{CN}(0, \sigma_n^2 \mathbf I)$.
	
	 By the relationship between the spatial-domain channel $\mathbf{h}_s$ and the angular-domain channel $\mathbf h_a$ in \eqref{beamspace_channel_vector}, the channel vector $\mathbf{h}_s$ in \eqref{received_pilots} can be replaced by $\mathbf h_s = \mathbf U^{H} \mathbf h_a$.
	 Thus, the received pilot symbols in \eqref{received_pilots} can be rewritten as
    \begin{IEEEeqnarray*}{lCl}
    \label{received_pilots_beamspace}
    \mathbf{r} =  \mathbf{P} \mathbf U^{H} \mathbf h_a + \mathbf{w}.
    \IEEEyesnumber
    \end{IEEEeqnarray*}
    By writing $\mathbf{\Phi} =\mathbf{P} \mathbf U^{H}$, 
    eq. \eqref{received_pilots_beamspace} can be expressed as 
     \begin{IEEEeqnarray*}{lCl}
     \label{CS_estimation}
    \mathbf{r} =\mathbf{\Phi} \mathbf h_a + \mathbf{w}
     \IEEEyesnumber  
    \end{IEEEeqnarray*}
     where $\mathbf \Phi \in \mathbb{C}^{M \times N}$. 
     We aim to estimate the sparse channel vector $\mathbf h_a  \in \mathbb{C}^{N}$ from the lower-dimensional measurements $\mathbf r  \in \mathbb{C}^{M}$ for $M \ll N$.
     Since the pilot length $M$ depends on the sparsity level of channel vector $\mathbf h_a$ instead of the number of BS antennas, the overheads of pilot transmission and CSI feedback can be largely reduced.
     
     In this paper, we adopt the Gaussian random measurement matrix $\mathbf{\Phi} \in \mathbb{R}^{M \times N}$, which has the real-form elements following the standard Gaussian distribution.  
     Thus, the real and imaginary parts of all the complex variables in \eqref{CS_estimation} can be written as 
     \begin{IEEEeqnarray*}{lCl}
	\label{CS_separate}
	\Re(\mathbf{r}) &= \mathbf{\Phi} \cdot \Re(\mathbf h_a) + \Re(\mathbf{w}) \\
	\Im(\mathbf{r}) &= \mathbf{\Phi} \cdot \Im(\mathbf h_a) + \Im(\mathbf{w})
	\IEEEyesnumber
	\end{IEEEeqnarray*}
	where $\Re(\cdot)$ and $\Im(\cdot)$ denote the real part and imaginary part of a complex vector. 
	Eq. \eqref{CS_separate} implies we can equivalently treat a complex channel vector $\mathbf h_a$ as a concatenated real-form vector 
	\begin{IEEEeqnarray*}{lCl}
	\label{concat_x}
	\mathbf{x} = [\Re(\mathbf h_a)^T, \Im(\mathbf h_a)^T]^T.
	\IEEEyesnumber
	\end{IEEEeqnarray*}
	In the remainder, we uniquely refer $\mathbf x$ as the sparse channel vector.

\section{DC Gradient Projection Sparse Reconstruction Algorithm}
\subsection{Exact Sparsity Constraint Representation}
	According to \eqref{CS_separate} and \eqref{concat_x}, the sparse channel estimation problem can be uniquely expressed using the real-form underdetermined linear system
	\begin{IEEEeqnarray*}{lCl}
	\label{CS_unified}
	\mathbf{y} =  \mathbf{\Phi} \mathbf{x} + \mathbf{n}
	\IEEEyesnumber
	\end{IEEEeqnarray*}
    where $\mathbf{y} \in \mathbb{R}^{M}$ denotes the compressed measurements;  $\mathbf{\Phi} \in \mathbb{R} ^ {M \times N}$ is the measurement matrix;
    $\mathbf{x} \in \mathbb{R}^{N}$ represents the sparse channel vector; 
    $\mathbf{n} \in \mathbb{R}^{M}$ is the Gaussian noise vector and $\mathbf{n} \sim \mathcal{N}(0, \sigma^2 \mathbf I)$.
	Sparse channel reconstruction is to obtain a solution of $\hat{\mathbf x}$ from compressed measurements $\mathbf y$ and measurement matrix $\mathbf \Phi$ such that $\hat{\mathbf x} \approx \mathbf x$.
	 This \mbox{$\ell_0$-minimization} problem is NP-hard and it is defined as
	\begin{IEEEeqnarray*}{lCl}
	\label{sparse_recovery1}
	\mathop{\text{min}}\limits_{\mathbf{x}} &&\quad \norm{\mathbf{x}}_0 \\
	\text{s.t.} &&\quad \norm{\mathbf{y}-\mathbf{\Phi}\mathbf{x}}_2^2 \le \tau
	\IEEEyesnumber
	\end{IEEEeqnarray*} 
	where $\tau$ is a nonnegative real-form parameter.
	Problem \eqref{sparse_recovery1} can be rewritten in an equivalent form as \cite{sparse2015rish}
	\begin{IEEEeqnarray*}{lCl}
	\label{sparse_recovery2}
	\mathop{\text{min}}\limits_{\mathbf{x}} &&\quad  \norm{\mathbf{y}-\mathbf{\Phi}\mathbf{x}}_2^2\\
	\text{s.t.} &&\quad \norm{\mathbf{x}}_0  \le K
	\IEEEyesnumber
	\end{IEEEeqnarray*} 
	where $K$ is the bound of the number of nonzero elements of vector $\mathbf x$, and it is uniquely determined by the parameter $\tau$ in problem \eqref{sparse_recovery1}.

	To represent exactly the sparsity constraint $\norm{\mathbf{x}}_0  \le K$ in \eqref{sparse_recovery2}, we introduce the top-$(K,1)$ norm.
	 The top-$(K,1)$ norm $\norm{\mathbf x}_{K,1}$ is defined as the sum of the largest $K$ elements of the vector $\mathbf x$ in terms of absolute value, namely 
	\begin{IEEEeqnarray*}{lCl}
	\norm{\mathbf x}_{K,1} :=  | x_{(1)} | +| x_{(2)} | + \cdots +| x_{(K)} |
	\IEEEyesnumber
	\end{IEEEeqnarray*} 
	where $| x_{(i)}|$ denotes the element whose absolute value is the $i$th-largest among the $N$ elements of the vector $\mathbf x$, i.e., $ | x_{(1)} | \ge| x_{(2)} | \ge \cdots \ge | x_{(N)} |$. 
	The constraint $\norm{\mathbf{x}}_0  \le K$ is equivalent to the statement that the $(K+1)$th-largest element of the vector $\mathbf x$ is zero, i.e., $\norm{\mathbf x}_{K+1, 1} - \norm{\mathbf x}_{K,1} = 0$. 
	Thus, we have an equivalent relationship between the following two statements \cite{Gotoh2018DC} 
	\begin{IEEEeqnarray*}{lCl}
	\norm{\mathbf{x}}_0  \le K \Leftrightarrow \norm{\mathbf x}_1 - \norm{\mathbf x}_{K,1}=0.
	\IEEEyesnumber
	\end{IEEEeqnarray*} 
	The problem \eqref{sparse_recovery2} can be rewritten as 
	\begin{IEEEeqnarray*}{lCl}
	\label{sparse_recovery3}
	\mathop{\text{min}}\limits_{\mathbf{x}} 
	&&\quad  \norm{\mathbf{y}-\mathbf{\Phi}\mathbf{x}}_2^2 \\
	\text{s.t.} 
	&&\quad \norm{\mathbf x}_1 - \norm{\mathbf x}_{K,1}=0
	\IEEEyesnumber
	\end{IEEEeqnarray*} 
	where the sparsity constraint $\norm{\mathbf x}_0 \le K$ is exactly represented by $\norm{\mathbf x}_1 - \norm{\mathbf x}_{K,1}= 0$.
	Using an appropriate Lagrange multiplier $\rho$, we rewrite the problem \eqref{sparse_recovery3} as the following unconstraint optimization problem
	 \begin{IEEEeqnarray*}{lCl}	
	 \label{dc_penalty_function}
	\mathop{\text{min}}\limits_{\mathbf{x}} \quad 
	\frac{1}{2}\norm{\mathbf{y}-\mathbf{\Phi}\mathbf{x}}_2^2 
	+ \rho (\norm{\mathbf x}_1 - \norm{\mathbf x}_{K,1})
	:=	F(\mathbf x) 
	\IEEEyesnumber
	\end{IEEEeqnarray*} 
	where $\rho$ is the regularization parameter that balance the data consistency and penalty term.
	Due to the nonnegativity of the penalty term $\rho (\norm{\mathbf x}_1 - \norm{\mathbf x}_{K,1}) \ge 0$, it can be ensured that the unconstrained problem \eqref{dc_penalty_function} is equivalent to the constraint problem \eqref{sparse_recovery3} when the penalty parameter $\rho$ is taking infinite limit, which can be proved in a similar way with Theorem 17.1 of \cite{Nocedal2006numerical}. 
	
	 To this end, we exactly represent the $\ell_0$-constraint $\norm{\mathbf x}_0 \le K$ using the DC constraint $\norm{\mathbf x}_1 - \norm{\mathbf x}_{K,1} = 0$ so that the sparse reconstruction problem \eqref{sparse_recovery2} is expressed in the equivalent form of \eqref{sparse_recovery3}.
	 Then we transform the problem \eqref{sparse_recovery3} into a unconstraint minimization problem \eqref{dc_penalty_function}.
	  The problem \eqref{dc_penalty_function} is a nonconvex optimization problem, because the subtracted \mbox{top-$(K,1)$} norm $\rho \norm{\mathbf x}_{K,1}$ in the penalty term results in a nonconvex regularizer $\rho (\norm{\mathbf x}_1 - \norm{\mathbf x}_{K,1})$.

\subsection{DC Programming Algorithm Framework}
	Our goal now is to solve the nonconvex unconstraint optimization problem \eqref{dc_penalty_function}.
	We employ the DC programming and decompose the objective function $F(\mathbf x)$ in problem \eqref{dc_penalty_function} as the difference of the two convex functions of $f(\mathbf x)$ and $g(\mathbf x)$ 
	\begin{IEEEeqnarray*}{lCl}
	\label{dc_quadratic_proj}
	\mathop {\text{min}}_\mathbf x 
	\underbrace{ \frac{1}{2}\norm{\mathbf{y}-\mathbf{\Phi}\mathbf{x}}_2^2  + \rho \norm {\mathbf x}_1}_{f(\mathbf x)}
	- \underbrace{ \rho \norm {\mathbf x}_{K,1}}_{g(\mathbf x)}.
	\IEEEyesnumber
	\end{IEEEeqnarray*}	 
	At the $t$th-iteration, we solve the following convex subproblem 
	\begin{IEEEeqnarray*}{lCl}
	\label{general_dc}
	\mathop{\text{min}}\limits_{\mathbf{x}} \quad  f(\mathbf x)
	- \mathbf x^T \partial g(\mathbf x^{t-1}).
	\IEEEyesnumber
	\end{IEEEeqnarray*}
	The second convex function $g(\mathbf x)$ in \eqref{dc_quadratic_proj} is linearized by $\mathbf x^T \partial g(\mathbf x^{t-1})$ in \eqref{general_dc}, where $\partial g(\mathbf x^{t-1})$ is the subgradient of $g(\mathbf x)$ at the $(t-1)$th update $\mathbf x^{t-1}$, that is
	\begin{IEEEeqnarray*}{lCl}
	\partial g(\mathbf x^{t-1}) =  
	 \rho \partial \norm{\mathbf x^{t-1}}_{K,1}
	\IEEEyesnumber
	\end{IEEEeqnarray*}	
	where $\partial \norm{\mathbf x^{t-1}}_{K,1}$ denotes the subgradient of $ \norm{\mathbf x^{t-1}}_{K,1}$. 
	The subgradient $\partial \norm{\mathbf x}_{K,1}$ of $\norm{\mathbf x}_{K,1}$ is defined as \cite{Gotoh2018DC}
	\begin{IEEEeqnarray*}{lCl}	
	 \partial \norm{\mathbf x}_{K,1}
	 := \mathop{\text{argmax}}_\mathbf w \left\lbrace \sum_{i=1} ^{N} x_i w_i \Big\vert \sum_{i=1}^N |w_i| =K,  w_i \in [-1, 1]\right\rbrace. \\
	\IEEEyesnumber
	\end{IEEEeqnarray*} 
	A subgradient $\mathbf w_x^{t-1} \in \partial \norm{\mathbf x^{t-1}}_{K,1}$ can be simply obtained by assigning the sign of the first $K$ largest elements of $|\mathbf{x}^{t-1}|$ to the corresponding elements of $\mathbf w_x^{t-1}$, i.e.,	$(\mathbf w_x)^{t-1}_{(i)}=\sign(x_{(i)}^{t-1})$, where the subscript $(i)$ represents the $i$th element of a vector, and setting the other elements of $\mathbf w_x^{t-1}$ to be zeros.
	
	The DC algorithm framework for sparse reconstruction is outlined as follows:
	\begin{itemize}
	\item[1.] \textbf{Start:} Given a starting point $\mathbf x ^0$, and a small threashold parameter $\epsilon>0$.
	\item[2.] \textbf{Repeat:} For $t=1,2,\ldots$ \\ 
					\hspace*{4mm} Select a subgradient $\mathbf w_x^{t-1} \in \partial \norm{\mathbf x^{t-1}}_{K,1}$;\\
					\hspace*{4mm} Solve the convex subproblem \eqref{general_dc}, i.e., $\mathop{\text{min}}\limits_{\mathbf{x}}  f(\mathbf x)
	- \rho \mathbf x^T \mathbf w_x^{t-1}$, and obtain $\mathbf x^t$;\\
					\hspace*{4mm} Increment $t$;
	\item[3.] \textbf{End:} Until terminate condition satisfies.				
	\end{itemize}

\subsection{DC Gradient Projection Algorithm for Sparse Reconstruction}
	Following the aforementioned DC algorithm framework to solve the problem \eqref{dc_penalty_function}, at the $t$th iteration we need to solve a nonsmooth convex subproblem \eqref{general_dc}, which is
	\begin{IEEEeqnarray*}{lCl}
	\label{subprob_quadratic_proj}
	\mathop{\text{min}}\limits_{\mathbf{x}} \quad 
	\frac{1}{2}\norm{\mathbf{y}-\mathbf{\Phi}\mathbf{x}}_2^2
	+ \rho \norm{\mathbf x}_1 
	- \rho \mathbf x^T \mathbf w_x^{t-1}
	\IEEEyesnumber
	\end{IEEEeqnarray*}
	where $\mathbf w_x^{t-1} = \partial \norm{\mathbf x^{t-1}}_{K,1}$.
		We turn it into a constraint quadratic problem and solve it using the projected gradient descent method.
	We split the positive and negative part of $\mathbf x$, and represent $\mathbf x$ as the difference of its positive part $\mathbf u$ and its negative part $\mathbf v$, that is
	\begin{IEEEeqnarray*}{lCl}
	\label{pos_neg_x}
	\mathbf x = \mathbf u - \mathbf v, \mathbf u \ge \mathbf 0, \mathbf v \ge \mathbf 0
	\IEEEyesnumber
	\end{IEEEeqnarray*}
	where $\mathbf u = (\mathbf x)_+, \mathbf v = (-\mathbf x)_+$, where $(\cdot)_+$ represents a nonnegative-clipper operation that retains the nonnegative elements and sets negative elements be zeros. More precisely, $(\mathbf x)_+$ represents for each element $x$ in vector $\mathbf x$ we take $(x)_+ = \max \{0, x\}$;  $(-\mathbf x)_+$ represents for each element $-x$ in vector $-\mathbf x$ we take $(-x)_+ = \max \{0, -x\}$.
	Noticing that $\norm{\mathbf x}_1 = \mathbf 1^T \mathbf u 	+  \mathbf 1^T \mathbf v$,
	the subproblem \eqref{subprob_quadratic_proj} can be written as a bound-constrained quadratic 
program (BCQP) 
	\begin{IEEEeqnarray*}{lCl}
	\label{bcqp}
	 \mathop{\text{min}}\limits_{\mathbf{u, v}} \quad 
	 && \frac{1}{2}\norm{\mathbf{y}-\mathbf{\Phi}(\mathbf{u} - \mathbf v)}_2^2 
	 	+ \rho \mathbf 1^T \mathbf u  	+ \rho \mathbf 1^T \mathbf v 
	  - \rho \mathbf u^T \mathbf w_u^{t-1} \\
	  && - \rho \mathbf v^T \mathbf w_v^{t-1} \\
	  \text{s.t.} \quad && \mathbf{u} \geq \mathbf{0}, \mathbf{v} \geq \mathbf{0}
	\IEEEyesnumber
	\end{IEEEeqnarray*}
	where $\mathbf w_u^{t-1}$ and $\mathbf w_v^{t-1}$ represent the positive and negative part of $\mathbf w_x^{t-1}$, i.e., $\mathbf w_u^{t-1} = (\mathbf w_x^{t-1})_+$, $\mathbf w_v^{t-1} = (-\mathbf w_x^{t-1})_+$.
	Let $\mathbf{z}$ denote the concatenation of $\mathbf{u}$ and $\mathbf{v}$, i.e., $\mathbf{z} = [\mathbf{u}^{T}, \mathbf{v}^{T}]^{T}$,
	we rewrite \eqref{bcqp} into a compact form
\begin{IEEEeqnarray*}{lCl}\label{compact_bcqp}
    \mathop{\text{min}}\limits_{\mathbf{z}} \quad 
    &&  \frac{1}{2} \mathbf{z}^{T} \mathbf{B} \mathbf{z} 
    + \mathbf c^T \mathbf z
    := G(\mathbf{z}) ,\\ 
    \text{s.t.} \quad && \mathbf{z} \geq \mathbf{0}
    \IEEEyesnumber
\end{IEEEeqnarray*}
where 
    \begin{IEEEeqnarray*}{lCl}
     \mathbf{z} = \left[\begin{matrix}
    \mathbf{u}\\
    \mathbf{v}
    \end{matrix}\right],
    \quad
    \mathbf{B} = \left[
    \begin{matrix}
    \mathbf{\Phi}^{T} \mathbf{\Phi} & -\mathbf{\Phi}^{T} \mathbf{\Phi}\\
    - \mathbf{\Phi}^{T} \mathbf{\Phi} & \mathbf{\Phi}^{T} \mathbf{\Phi}
    \end{matrix}
    \right],
    \quad \\
    \mathbf c = \left[\begin{matrix}
    \mathbf{\Phi}^{T} \mathbf{y}\\
    -\mathbf{\Phi}^{T} \mathbf{y}
    \end{matrix}\right]
    +\rho \mathbf 1^T
    - \rho \mathbf w_z^{t-1}
	\end{IEEEeqnarray*}
where $\mathbf 1^T$ represents the all-ones column vector in the same size with $\mathbf z$;    
$\mathbf s_z^{t-1} = [(\mathbf s_u^{t-1})^T, (\mathbf s_v^{t-1})^T]^T \in \partial \norm{ \mathbf  z^{t-1}}_{K, 1}$ is a subgradient of $\norm{ \mathbf  z^{t-1}}_{K, 1}$, and $\mathbf w_z^{t-1} $ can have either zero-valued or one-valued elements for $\mathbf z^{t-1} \ge 0$.

	Now, we can apply the gradient projection descent to solve the subproblem \eqref{compact_bcqp}. 
	The $k$th-step update is expressed as
    \begin{IEEEeqnarray*}{lCl}
    \label{accelerated_pgd_general}
    \mathbf{z}^{(k+\frac{1}{2})} &= \textit{Proj} \left(\mathbf{z}^{(k)} - \alpha^k \nabla G(\mathbf{z}^{(k)})\right),\\
     \mathbf{z}^{(k+1)} &= \mathbf{z}^{(k)} +\beta^{k} ( \mathbf{z}^{(k+\frac{1}{2})}  -  \mathbf{z}^{(k)} )
     \IEEEyesnumber
    \end{IEEEeqnarray*}
    where $\alpha^{k} > 0$ is the step size which can be determined by the Barzilai-Borwein (BB) approach, and $\beta^{k} \in (0, 1]$ is another step size can be found by line search \cite{figueiredo2007gradient};
    $\textit{Proj}(\cdot)$ represents the operation of orthogonal projection that projects the vector to the nonnegative orthant;
     $\nabla G(\mathbf{z}^{(k)})$ represents the gradient of $G(\mathbf z)$ in terms of $\mathbf{z}^{(k)}$ which is calculated as
    \begin{IEEEeqnarray*}{lCl}
    \label{gradient_Gz}
   \nabla G(\mathbf{z}^{(k)}) 
  	&&=  
   \left[
    \begin{matrix}
    \mathbf{\Phi}^{T} \mathbf{\Phi} (\mathbf{u}^{(k)}-\mathbf{v}^{(k)})\\
    -\mathbf{\Phi}^{T} \mathbf{\Phi} (\mathbf{u}^{(k)}-\mathbf{v}^{(k)})
     \end{matrix}
    \right]
    + \left[\begin{matrix}
    -\mathbf{\Phi}^{T} \mathbf{y}\\
    \mathbf{\Phi}^{T} \mathbf{y}
    \end{matrix}\right] \\
    && - \rho \mathbf w_z ^{t-1}  +\rho \mathbf{1}^T. 
    \IEEEyesnumber
    \end{IEEEeqnarray*}

	In summary, the DC algorithm can be simplified as iteratively performing the following two steps until convergence:
	\begin{IEEEeqnarray*}{lCl}
	\label{dc_projection}
	(\text a) 
	&& \quad \mathbf w_z^{t-1} 
     \in  \partial \norm{\mathbf z^{t-1}}_{K, 1}\\
	(\text {b}) 
	&& \quad \mathbf z^{t} 
	= \mathop{\text{argmin}}_{\mathbf z \ge 0} 
	\left\lbrace 
	\frac{1}{2} \mathbf{z}^{T} \mathbf{B} \mathbf{z} + \mathbf c^T \mathbf z	
    \right\rbrace 
	\IEEEyesnumber  
	\end{IEEEeqnarray*} 
    where $\mathbf{B}$ and $\mathbf c$ are defined in \eqref{compact_bcqp}.
    The subproblem (b) in \eqref{dc_projection} is solved by applying the gradient projection descent \eqref{accelerated_pgd_general}. We summarize this DC gradient projection algorithm for sparse reconstruction (DC-GPSR) in Algorithm 1.

\begin{algorithm}
	\caption{DC gradient projection descent algorithm for sparse reconstruction (DC-GPSR)} 
	\hspace*{\algorithmicindent} \textbf{Input:} measurements $\mathbf y$, measurement matrix $\mathbf \Phi$ and a small number $\epsilon$ \\
    \hspace*{\algorithmicindent} \textbf{Output:} reconstructed $\hat{\mathbf x}$ \\
    	\hspace*{\algorithmicindent} \textbf{Initialization:}  $\mathbf u^0$, $\mathbf v^0$, $\mathbf z^0 \leftarrow [(\mathbf u^0)^T,  (\mathbf v^0)^T]^T$
	\begin{algorithmic}[1]
		\For {$t=1,2,\ldots$}
			\State Compute the subgradient $\mathbf w_z^{t-1}  \in \partial \norm{\mathbf z^{t-1}}_{K,1}$ 
			\For {$k=1,2,\ldots$}
				\State Compute gradient $\nabla G(\mathbf{z}^{(k)})$ by \eqref{gradient_Gz}
				\State Perform gradient projection descent \eqref{accelerated_pgd_general} for obtaining $\mathbf z^{(k+1)}$
				\State Check convergence, set $\mathbf z^* \leftarrow \mathbf z^{(k+1)}$ and proceed to step 7 if convergence is satisfied; otherwise return to step 3.
			\EndFor
			\State $\mathbf z^t \leftarrow \mathbf z^*$
			\State Check the terminate condition $\norm{\mathbf z^{t}-\mathbf z^{t-1}}_2 \le \epsilon$, and return to step 1 if not satisfied; otherwise terminate with approximate solution $\mathbf z^t =  [(\mathbf u^t)^T,  (\mathbf v^t)^T]^T$, and obtain the reconstruction $\hat{\mathbf x} = \mathbf u^t - \mathbf v^t$.
		\EndFor
	\end{algorithmic} 
\end{algorithm}

\section{Numerical Result}
	In this section, the performance of the proposed DC-GPSR algorithm is numerically evaluated and compared with existing recovery algorithms.
	In our simulation, the number of BS antennas is set as $N =256$; 
	the sparsity level of the angular-domain channel is set as $16$;
	the dimension of measurements is set as $M = 128$.
	The real and imaginary part of the channel vector is concatenated together being a sample of sparse channel vector. 
	Thus, we are reconstructing a sparse vector $\hat{\mathbf x} \in \mathbb{R}^{512}$ with a number of nonzero elements $K=32$ from compressed measurements $\mathbf y \in \mathbb{R}^{128}$.
	A random Gaussian matrix drawn from standard Gaussian distribution is adopted as the measurement matrix $\mathbf \Phi$;
	The terminate condition for our proposed algorithm is set as $\norm{\hat{\mathbf x}^{t}-\hat{\mathbf x}^{t-1}}_2 \le 10^{-30}$.

\subsection{Performance in a Noiseless Scenario}
	
\begin{figure}[t]
\centering
\normalsize
\includegraphics[width=3.2in]{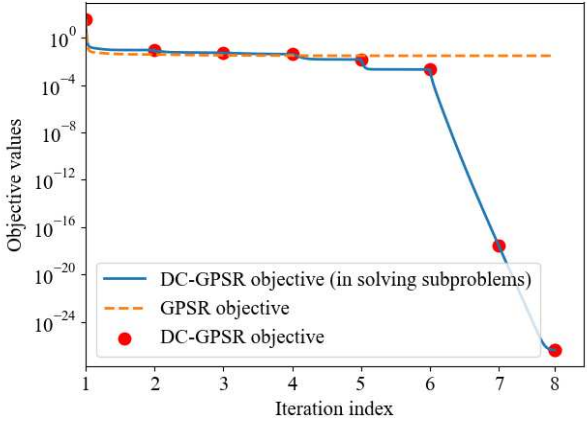}
\captionsetup{justification=centering}
\caption{Objective values versus iterations; DC-GPSR objective is $\frac{1}{2}\norm{\mathbf{y}-\mathbf{\Phi}\hat{\mathbf{x}}}_2^2 
	+ \rho (\norm{\hat{\mathbf x}}_1 - \norm{\hat{\mathbf x}}_{K,1})$, and GPSR objective is $\frac{1}{2}\norm{\mathbf{y}-\mathbf{\Phi}\hat{\mathbf{x}}}_2^2 + \rho \norm{\hat{\mathbf x}}_1$}
\label{fig1}
\end{figure}

\begin{figure}[t]
\centering
\normalsize
\includegraphics[width=3.2in]{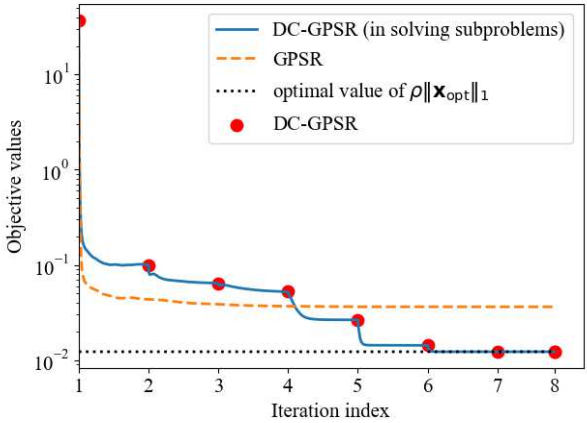}
\captionsetup{justification=centering}
\caption{Objective values ($\frac{1}{2}\norm{\mathbf{y}-\mathbf{\Phi}\hat{\mathbf{x}}}_2^2 + \rho \norm{\hat{\mathbf x}}_1$) versus iterations for DC-GPSR and GPSR algorithm}
\label{fig2}
\end{figure}

\begin{figure}[t]
\centering
\normalsize
\includegraphics[width=3.2in]{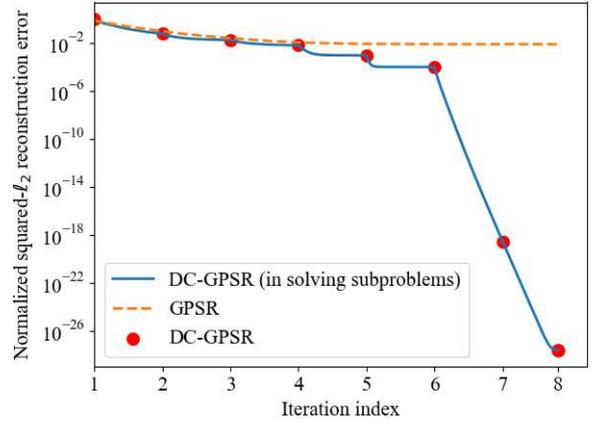}
\captionsetup{justification=centering}
\caption{Normalized squared-$\ell_2$-error of reconstructions versus iterations for DC-GPSR and GPSR algorithm}
\label{fig3}
\end{figure}

\begin{figure}[t]
\centering
\normalsize
\includegraphics[width=3.3in]{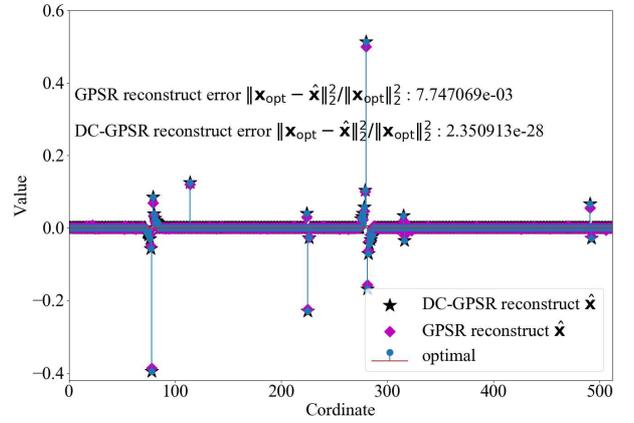}
\captionsetup{justification=centering}
\caption{Illustration of the optimal sample and the final reconstructions by DC-GPSR and GPSR algorithm}
\label{fig4}
\end{figure}

	We perform the proposed DC-GPSR algorithm in \mbox{noiseless} scenario to reconstruct a channel vector.
	The optimal sparse channel vector is denoted by $\mathbf x_{\text{opt}}$; and the reconstructed vector is denoted by $\hat{\mathbf x}$.
	The CPU time of running DC-GPSR algorithm is $0.086$ seconds on a desktop computer equipped with 3.2 GHz Intel Core i7-8700 CPU.
	The objective evolutions versus iterations are shown in Fig. \ref{fig1} and Fig. \ref{fig2}.
	The reconstruction errors versus iterations is shown in Fig. \ref{fig3}.
	The optimal and reconstructed vectors are plotted in Fig. \ref{fig4}.
	We plot the results of conventional GPSR algorithm \cite{figueiredo2007gradient} in Figs. \ref{fig1}--\ref{fig4} as comparisons.

	Figure \ref{fig1} shows the objective values versus iterations, where the red dots indicate iterations of the DC-GPSR algorithm, and the solid line shows the objective evolutions of gradient projection for solving the subproblem at each iteration.
	The dashed line shows the objective of conventional GPSR algorithm for comparison.
	We can see the proposed DC-GPSR algorithm quickly decreases its objective after the sixth \mbox{outer-step}, and terminates at the the eighth step, while the objective of GPSR is stuck at a relatively large value.
	
	We investigate the values of least square objective with the $\ell_1$-norm penalty, i.e., $\norm{\mathbf y - \mathbf \Phi \hat{\mathbf x}}_2^2 + \rho \norm{\hat{\mathbf x}}_1$, and show the values versus iterations in \mbox{Fig. \ref{fig2}}. 
	The solid line marked by the red dots indicates the result of DC-GPSR, and the dashed line indicates the result of GPSR. 
	We also draw the optimal values of the penalty term $\rho \norm{\mathbf x_{\text{opt}}}_1$.
	We can see the \mbox{DC-GPSR} algorithm arrives and stays at the value of $\rho \norm{\mathbf x_{\text{opt}}}_1$ after the sixth \mbox{iteration}, which is also the optimal value that the objective $\norm{\mathbf y - \mathbf \Phi \hat{\mathbf x}}_2^2 + \rho \norm{\hat{\mathbf x}}_1$ can achieve when $\norm{\mathbf y - \mathbf \Phi \hat{\mathbf x}}_2^2 = 0$ and $\rho \norm{\hat{\mathbf x}}_1 = \rho \norm{\mathbf x_{\text{opt}}}_1$. 
	However, the GPSR cannot achieve this optimal value with a gap.
	It is meaningful to observe this is the gap of minimal objective values between GPSR and DC-GPSR, because it help us understand the approximation error from relaxing the $\ell_0$-norm by the $\ell_1$-norm.

	\mbox{Figure \ref{fig3}} shows the normalized squared-$\ell_2$ error (\mbox{$\norm{\mathbf x_\text{opt}-\hat{\mathbf x}}_2^2 / \norm{\mathbf x_\text{opt}}_2^2$}) versus iterations for reconstructions by DC-GPSR and GPSR algorithm.
	\mbox{Fig. \ref{fig4}} illustrates the optimal sample and the finally reconstructed vectors by DC-GPSR and GPSR algorithm.
	We can see the proposed DC-GPSR algorithm achieves an accurate reconstruction with a small error on the order of $10^{-28}$, which is far more accurate than the reconstruction by the GPSR algorithm with error on the order of $10^{-3}$.

\subsection{Performance in a Noisy Scenario}

\begin{figure}[t]
\centering
\normalsize
\includegraphics[width=3.2in]{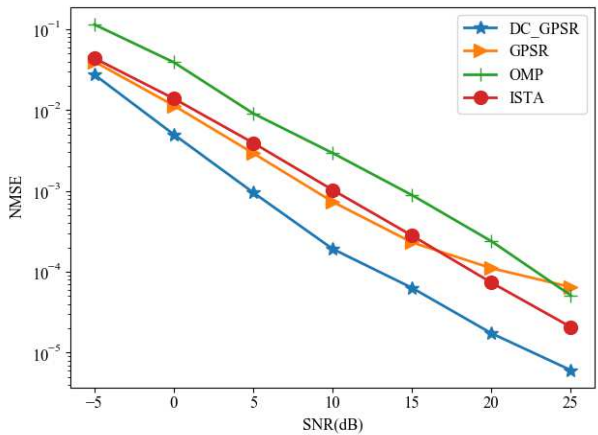}
\captionsetup{justification=centering}
\caption{Reconstruction NMSE in noisy scenarios by different sparse reconstruction algorithms}
\label{fig6}
\end{figure}

	We perform sparse reconstructions by DC-GPSR in the noisy scenarios for $1,000$ channel vector samples, and compare the normalized mean square error (NMSE) with several existing reconstruction algorithms. 
	The NMSE is defined as 
\begin{IEEEeqnarray*}{lCl}
\text{NMSE} = \frac{1}{m} \sum_{i=0}^{m-1}  \frac{\norm{\mathbf x_i- \hat{\mathbf x}_i}_2^2}{\norm{\mathbf x_i}_2^2}
\end{IEEEeqnarray*}
	where $m=1000$ is the number of samples. 
	For the $i$th channel vector sample $\mathbf x_i$, we add noise $\mathbf n_i \in \mathbb{R}^{128}$ on the compressed measurements $\mathbf y_i$ to form a noise-corrupted compressed measurements $\mathbf y_i^{\prime}$, then we reconstruct the sparse vector $\hat{\mathbf x_i}$ from $\mathbf y_i^{\prime}$ and measurement matrix $\mathbf \Phi$.
	The signal-to-noise ratio (SNR) is defined as 
\begin{IEEEeqnarray*}{lCl}
\text{SNR}  = \frac{\norm{\mathbf x_i}_2^2}{\mathbb{E}(\norm{\mathbf n_i}_2^2)} =\frac{\norm{\mathbf x_i}_2^2}{M\cdot \sigma^2}
\end{IEEEeqnarray*}
where the subscript $i=1,2,...,1000$ is the sample index; $\mathbb{E}(\cdot)$ is the expectation operator; $M$ is the dimension of compressed measurements; $\sigma$ is the standard deviation of the noise that follows zero-mean Gaussian distribution.

	The benchmark algorithms include ISTA \cite{blumensath2008thomas}, GPSR \cite{figueiredo2007gradient} and OMP \cite{tropp2007signal}. 
	The results of reconstruction NMSE are shown in Fig. \ref{fig6}.	
	We can see the proposed DC-GPSR has a lower reconstruction NMSE compared with the existing sparse reconstruction algorithms.
	Although DC-GPSR performance degrades due to noise, the DC-GPSR algorithm exhibits considerable robustness in the noisy scenario.
	For example, when the SNR is $25$ dB, the reconstruction NMSE of DC-GPSR algorithm is $6.07\times 10^{-6}$, which is sufficiently accurate for most of the applications.

\section{Conclusion}

	A novel sparse reconstruction algorithm DC-GPSR was proposed for the sparse channel estimation in massive MIMO systems.
	The sparse recover problem was formulated as a least squares problem with a nonconvex regularizer.
	The nonconvex regularizer is an exact representation for the \mbox{$\ell_0$-norm} constraint, which leads to more accurate sparse solution than the convex \mbox{$\ell_1$-norm} regularizer.
	In the proposed DC-GPSR algorithm, the nonconvex optimization problem is decomposed and approximated by a list of convex subproblems; the convex subproblems are expressed as BCQP problems and solved by the gradient projection descent method.
	The proposed DC-GPSR algorithm shares the global convergence property with the general DC algorithm.
	Numerical results showed the DC-GPSR algorithm is fast, accurate and robust, and it outperforms several existing sparse recovery algorithms.

\bibliographystyle{IEEEtran}

\bibliography{IEEEabrv,mybib}

\end{document}